\documentclass[twocolumn,aps,10pt,prd]{revtex4-2}

\usepackage[dvips]{graphicx}
\usepackage{epsfig,amsmath,amssymb,verbatim,mathrsfs,array,layout,textcomp,latexsym, slashed,graphicx,bbm}

\usepackage[normalem]{ulem}
\usepackage{appendix}

\usepackage{rotating}
\usepackage{hyperref}
\usepackage{xcolor}
\usepackage{soul}

\newcommand{\be}{\begin{equation}}
\newcommand{\ee}{\end{equation}}
\newcommand{\bea}{\begin{eqnarray}}
\newcommand{\eea}{\end{eqnarray}}

\definecolor{gbcolor}{rgb}{.8,.3,.1}
\definecolor{gbcolor2}{rgb}{.8,.1,.7}

\def\beq{\begin{equation}}
\def\eeq{\end{equation}}

\begin{document}

\begin{flushleft}                          
\footnotesize DESY 22-037,  TUM-HEP-1389-22
\end{flushleft}

\title{Revealing the Cosmic History with Gravitational Waves}

\author{Andreas Ringwald$^{1}$}\email{andreas.ringwald@desy.de}
\author{Carlos Tamarit$^2$}\email{carlos.tamarit@tum.de}

\affiliation{$^1$Deutsches Elektronen-Synchrotron DESY, Notkestr. 85, 22607 Hamburg, Germany}
\affiliation{$^2$ Physik-Department T70, Technische Universit\"at M\"unchen, James-Franck-Stra\ss e, 85748 Garching, Germany}

\begin{abstract}

The characteristics of the cosmic microwave background provide circumstantial evidence that the hot radiation-dominated epoch in the early universe was preceded by a period of inflationary expansion. Here, we show how a measurement of the stochastic gravitational wave background can reveal the cosmic history and the physical conditions during inflation, subsequent pre- and reheating, and the beginning of the hot big bang era. This is exemplified with a particularly well-motivated and predictive minimal extension of the 
Standard Model  which is known to provide a complete model for particle physics --up to the Planck scale-- and for cosmology --back to inflation.

\end{abstract}

\maketitle

\section{Introduction.} Big Bang cosmology describes how the universe expanded from an initial state of extremely high density 
into the cosmos we currently inhabit. It comprehensively explains a broad range of observed phenomena, including the abundance of light elements, the Cosmic Microwave Background (CMB) radiation, and the large-scale structure.   
It successfully delineates the cosmic history back to at least a fraction of a second after its birth, 
when the primordial plasma was radiation-dominated and Big Bang Nucleosynthesis (BBN) took place, 
at temperatures around a few MeV. 
 
Direct information about the cosmic history prior to BBN may be obtained from the observation of Gravitational Waves (GWs). 
In fact, after their production they freely traverse cosmic distances, making them a unique probe of the very early universe~\cite{Maggiore:2018sht,Caprini:2018mtu}.  An eventual measurement of the complete spectrum of primordial stochastic GWs may inform us in particular about three cosmological events 
supposed to have occurred in cosmic history: {\em i)} a 
stage of inflationary expansion preceding the radiation-dominated era, {\em ii)} the subsequent pre- and reheating stages, and {\em iii)} the 
beginning of the hot thermal radiation-dominated era after reheating.

The corresponding GW predictions are model-dependent. 
They depend crucially on the field content and its dynamics, in particular on the parameters determining the scale of inflation and the reheating temperature. 
To get the complete picture, one needs a complete model for particle physics and cosmology, such as for example
the Standard Model*Axion*Seesaw*Higgs portal inflation (SMASH) model~\cite{Ballesteros:2016euj,Ballesteros:2016xej,Ballesteros:2019tvf} -- a well motivated and predictive minimal extension of the Standard Model of particle physics (SM) 
which addresses five fundamental problems of particle physics and cosmology  in one stroke: inflation, baryon asymmetry, neutrino masses, strong CP problem, and dark matter. 

\begin{figure}[h]
\begin{center}
\includegraphics[width=.87\columnwidth]{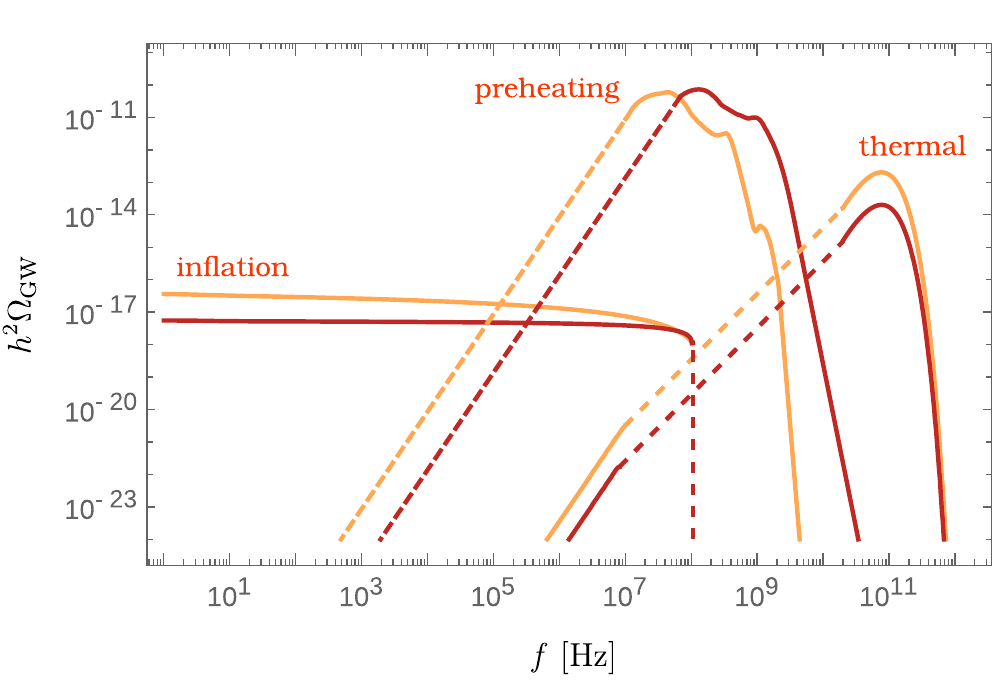}
\end{center}
\vskip-.5cm 
\caption{Today's fractional contribution of primordial GWs to the energy density in the universe per logarithmic frequency interval, $h^2 \Omega_{\rm GW}$, versus the frequency, $f$, as predicted in  SMASH  for the benchmark points 1 (lighter) and 2 (darker). 
}
\label{fig:h2Omega_smash_entire}
\end{figure}

{In SMASH, once the model parameters are fixed, the spectrum of stochastic GWs is calculable. As such, the contributions from different physical processes are not independent and their features will be correlated. The stochastic GW spectrum in SMASH receives contributions from quantum fluctuations during inflation, inflaton fragmentation during preheating \cite{Khlebnikov:1997di,Easther:2006gt,Easther:2006vd,Dufaux:2007pt,Garcia-Bellido:2007nns,Garcia-Bellido:2007fiu,Easther:2007vj,Dufaux:2008dn}, and thermal fluctuations  at the beginning of the hot thermal radiation-dominated stage\cite{Ghiglieri:2015nfa,Ghiglieri:2020mhm,Ringwald:2020ist}. {The three sources are inter-dependent as each process determines the initial conditions for the subsequent one.} A hypothetical detection of the spectrum in different frequency ranges would allow to cross-check for the correlations predicted in SMASH, opening new possibilities for falsifying the model. As the latter features no sizable GW production from  sources such as cosmic strings or first-order phase transitions, the resulting spectrum can be seen as a conservative benchmark for high-frequency GW searches.}
In two preceding publications we have determined the GW spectra in SMASH originating  during 
inflation~\cite{Ringwald:2020vei} and from thermal fluctuations~\cite{Ringwald:2020ist}. In this paper we determine the GW spectrum arising during  preheating~ and, {using the results of the preheating simulations, we} provide improved estimates of the reheating temperature (first estimated in \cite{Ballesteros:2016xej}) and the ensuing spectrum of GWs from the thermal plasma, {which allows us to go beyond the estimates of Ref.~\cite{Ringwald:2020ist}}. To the best of our knowledge, this represents the first computation of the complete spectrum of stochastic GWs generated in the early universe for a particular particle physics model~\footnote{Though there have been other efforts to estimate complete primordial spectra, such as Ref.~\cite{Buchmuller:2013lra}, the GWs sourced by thermal fluctuations were not accounted for. Other computations such as \cite{Klose:2022knn} focused only on the thermal part.}, cf. Fig.~\ref{fig:h2Omega_smash_entire}. 

\section{The SMASH model.} In the SMASH model \cite{Ballesteros:2016euj,Ballesteros:2016xej,Ballesteros:2019tvf}, a new complex scalar field  $\sigma$ (the Peccei-Quinn (PQ) field),
a vectorlike quark $Q$ and three singlet neutrinos $N_i$, with $i=1,2,3$, are added to the SM. All the new fields, as well as the quarks and leptons of the SM, are assumed to be charged under a global $U(1)_{\rm PQ}$ symmetry. The scalar potential in SMASH, which involves also the Higgs doublet $H$ (neutral under PQ), has the general form: 
\begin{equation}
\label{scalar_potential}\begin{aligned}
V(H,\sigma )= &\lambda_H \left( H^\dagger H - \frac{v^2}{2}\right)^2
+\lambda_\sigma \left( |\sigma |^2 - \frac{v_{\sigma}^2}{2}\right)^2\\
&+
2\lambda_{H\sigma} \left( H^\dagger H - \frac{v^2}{2}\right) \left( |\sigma |^2 - \frac{v_{\sigma}^2}{2}\right)\,.
\end{aligned}
\end{equation}
Here, the dimensionless couplings are assumed to obey $\lambda_H, \lambda_\sigma >0$,  
$\lambda_{H\sigma}^2 <  \lambda_H \lambda_\sigma$, in order to ensure that the PQ and electroweak   symmetries are broken by the vacuum expectation
values 
$\langle H^\dagger H\rangle = v^2/2$, 
$\langle |\sigma |^2\rangle=v_{\sigma}^2/2$,
where $v_\sigma\gg v=246$\,GeV. The hypercharge of the vectorlike quark $Q$ and the PQ charges of the SM fermions are chosen such that the only allowed  interactions of the exotic fermions $N_i,Q$ are 
$ {\cal L}\supset 
-[F_{ij}\bar{ N}_j P_L L_i\epsilon H+\frac{1}{2}Y_{ij}\sigma \bar N_i P_L  N_j 
+y\, \sigma \bar Q P_L Q+\,{y_{Q_d}}_{i}\sigma\bar{D}_iP_L Q +h.c.]$.
In the previous formula the fermion fields are four-component spinors. $D_i,L_i$ denote the Dirac spinors associated with the down quarks and leptons of the $i$th generation, while the $N_i$ are taken to be Majorana spinors.
In this model the strong CP problem is solved by the PQ mechanism~\cite{Peccei:1977hh}. 
The axion~\cite{Weinberg:1977ma,Wilczek:1977pj} -- the pseudo Goldstone boson associated with the spontaneous breaking of the PQ symmetry --   can be the main constituent of dark matter if its decay constant 
$f_a = v_\sigma$ {between $\sim10^{10}$ GeV and} $\sim 10^{11}\,\mathrm{GeV}$~\cite{Preskill:1982cy,Abbott:1982af,Dine:1982ah}.
The PQ symmetry breaking scale also gives rise to large Majorana masses for the heavy neutrinos. This can explain the smallness of the active neutrinos' masses  through the seesaw mechanism~\cite{Minkowski:1977sc,Gell-Mann:1979vob,Yanagida:1979as,Mohapatra:1979ia} and also results in the generation of the baryon asymmetry of the universe via thermal leptogenesis~\cite{Fukugita:1986hr}. {Additionally, the instability of the Higgs potential at large field values, present for the preferred value of the top mass \cite{Degrassi:2012ry}, can be cured in SMASH by the stabilizing effect of the portal coupling $\lambda_{H\sigma}$ \cite{Lebedev:2012zw,Elias-Miro:2012eoi}. For $\lambda_{H\sigma}<0$, as necessary for a successful reheating, this requires $\lambda^2_{H\sigma}/\lambda_\sigma$ to be between $\sim 10^{-2}$ and $\sim10^{-1}$ \cite{Ballesteros:2016xej}. While higher values are allowed, they typically result in running couplings that become nonperturbative at large scales, with the ensuing loss of predictive power}.

\section{The cosmic history in SMASH.} 
\begin{figure}[t]
\begin{center}
\includegraphics[width=.95\columnwidth]{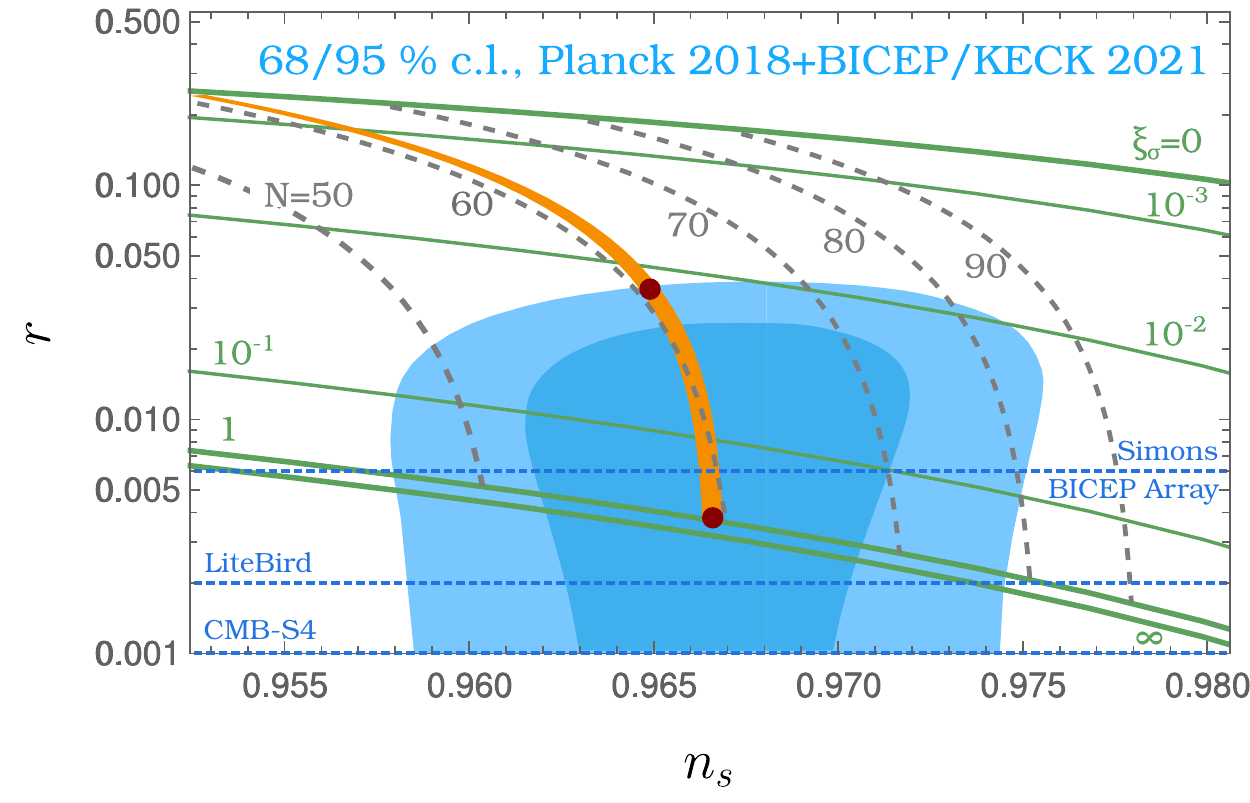}
\vskip-.2cm 
\caption{\label{fig:r_vs_ns} Inflationary predictions in SMASH
in the $r$ vs $n_{s}$ plane with a pivot scale of 0.002 Mpc$^{-1}$.
The green solid/dashed gray lines are contours of constant 
$\xi_\sigma$/number of efolds, respectively. Accounting for a consistent reheating history gives the orange region, and the red dots correspond to the benchmark scenarios BP1 (upper dot) and BP2 (lower dot). We also show the 68\% and 95\% C.L. contours arising from Planck and BICEP/KECK data \cite{BICEP:2021xfz}, {as well as the 95\% projected sensitivities from the Simons observatory \cite{SimonsObservatory:2018koc}, BICEP Array \cite{BICEP:2021xfz}, LiteBird \cite{LiteBIRD:2022cnt},  and  CMB-S4 \cite{Abazajian:2019eic}.}}
\end{center}
\end{figure}

\begin{figure}[t]
\begin{center}
\includegraphics[width=.95\columnwidth]{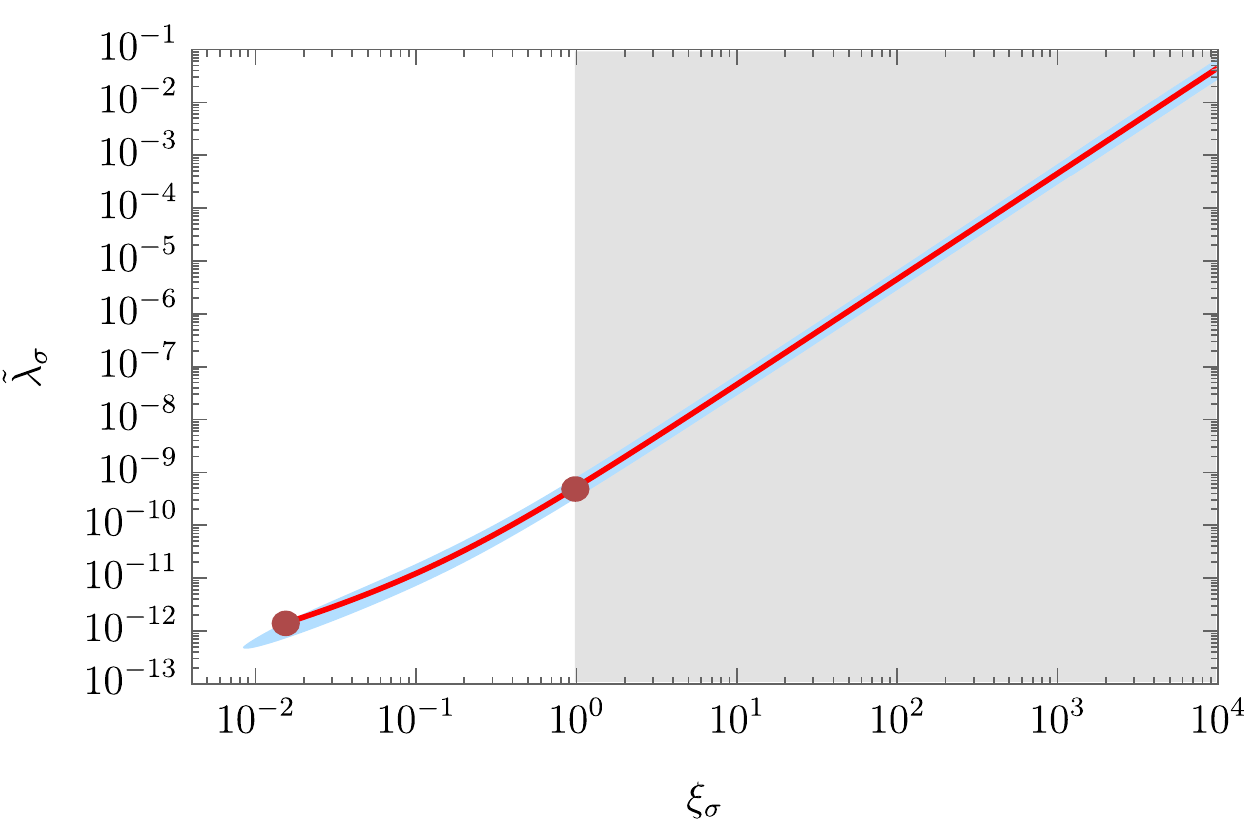}
\includegraphics[width=.95\columnwidth]{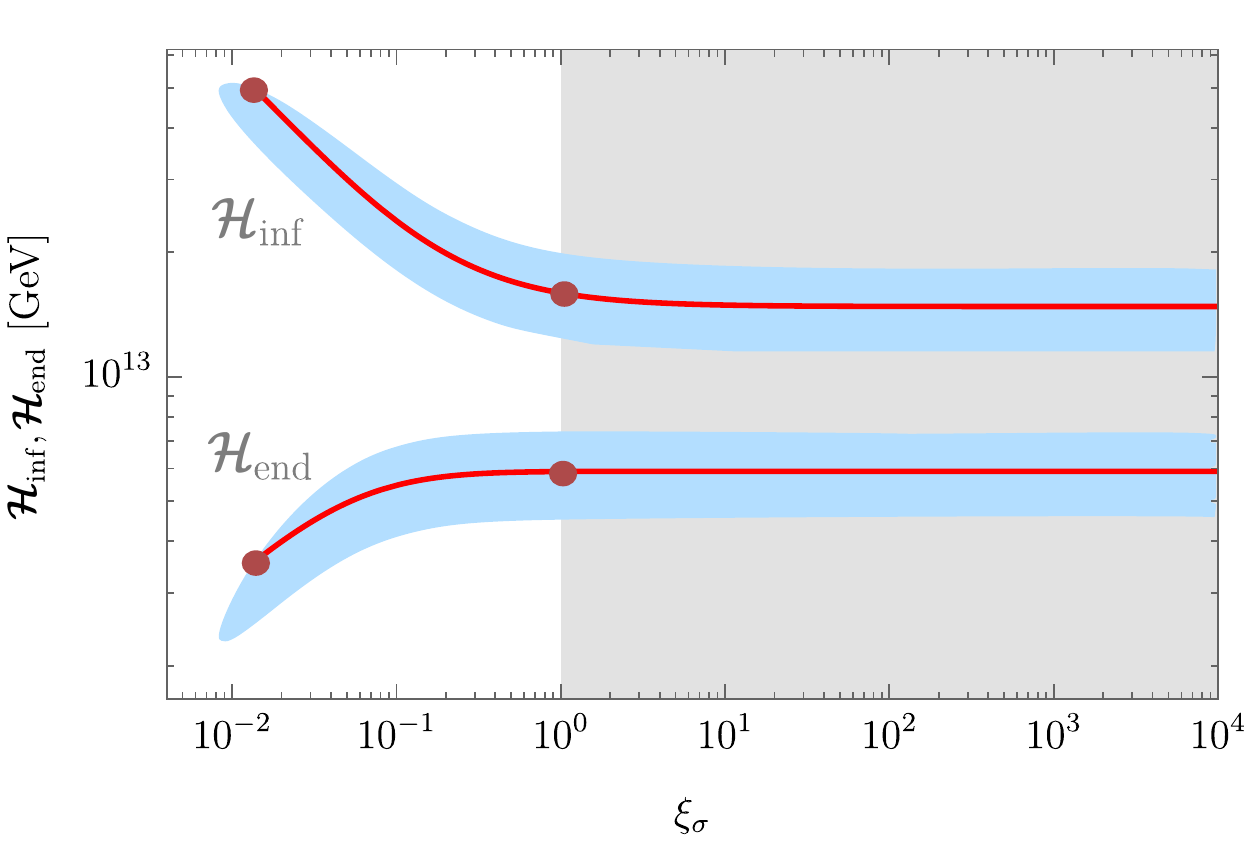}
\includegraphics[width=.95\columnwidth]{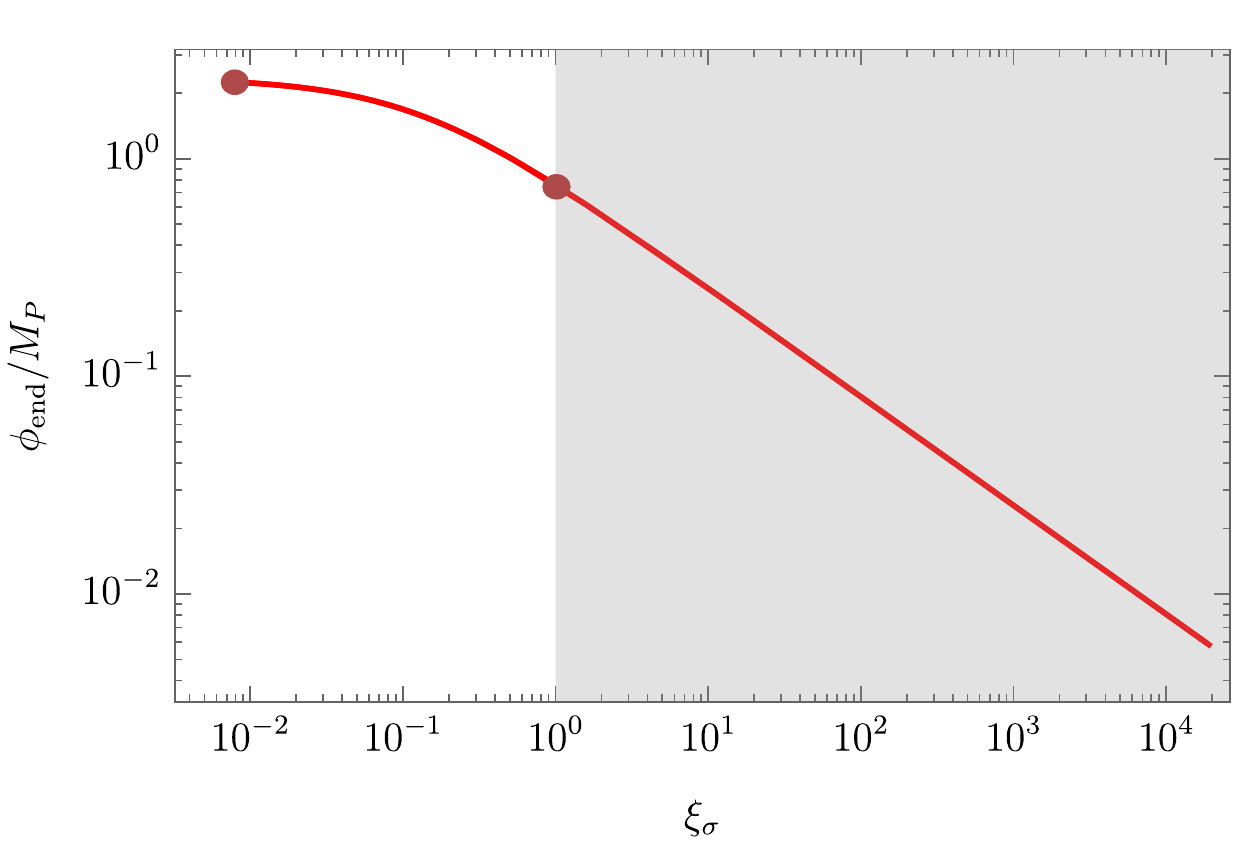}
\vskip-.2cm 
\caption{\label{fig:x_vs_xi}{Inflationary constraints/predictions on $\tilde\lambda_\sigma$ (top), Hubble scale at the beginning and end of inflation (middle) and field value at the end of inflation (bottom) as a function of $\xi_\sigma$. The red lines are the SMASH results, shown within the blue regions compatible with the 95\% C.L. contours of the latest  combination of Planck and BICEP/KECK data. The dots correspond to the benchmark points BP1 and BP2.}}
\end{center}
\end{figure}

Inflation results from the dynamics of the PQ  and Higgs fields in the presence of non-minimal couplings to the Ricci 
scalar $R$~\cite{Spokoiny:1984bd,Futamase:1987ua,Salopek:1988qh,Fakir:1990eg,Bezrukov:2007ep},
\begin{equation}
  \label{Lmain}
  S\supset - \int d^4x\sqrt{- g}\,\left[
     \frac{M^2}{2}  + \xi_H\, H^\dagger H+\xi_\sigma\, \sigma^* \sigma  
  \right] R
  \,.
\end{equation}
Here, the mass scale $M$ is related to the reduced Planck mass ($M_P\simeq 2.435\times 10^{18}\,\rm GeV$) by  
$M^2_P=M^2+\xi_H v^2+\xi_\sigma v^2_\sigma$. After a Weyl transformation of the metric to the Einstein frame, which eliminates the non-minimal gravitational couplings, the potential  becomes flat for large field values.
Problems with perturbative unitarity~\cite{Barbon:2009ya,Burgess:2009ea} are avoided by requiring $1\gtrsim \xi_\sigma\gg \xi_H\geq 0$; we will neglect $\xi_H$ in the following. To ensure a viable reheating scenario, slow-roll inflation should take place along an inflationary valley that can be approximated by the line $h/({\sqrt{2}\,{\rm Re}\sigma})=\sqrt{-\lambda_{H\sigma}/\lambda_H}$, where $h$  denotes the neutral component of the Higgs doublet in the unitary gauge. This requires a negative portal coupling   $\lambda_{H\sigma}<0$. {For positive $\lambda_{H\sigma}$, inflation can take place along the direction of e.g. ${\rm Re}\,\sigma$, but in this case reheating can be shown to be problematic, leading to an excess of dark radiation \cite{Ballesteros:2016xej}. Returning to $\lambda_{H\sigma}<0$, } he potential along the valley is determined {by two parameters: an effective coupling $\tilde\lambda_\sigma=\lambda_\sigma-\lambda_{H\sigma}^2/\lambda_H$, and $\xi_\sigma$}. With the power spectrum of scalar/tensor  perturbations during inflation parameterized as 
$
\Delta^2_{s/t}(k)=A_{s/t}(k_*)\left({k}/{k_*}\right)^{n_{s/t}(k_*)-1+\cdots}$, 
where $k_*$ is a given reference pivot scale, 
the predictions for the spectral index $n_s(k_*)$ and the tensor-to-scalar ratio 
$
 r={A_t(k_*)}/{A_s(k_*)}
$
are shown in Fig.~\ref{fig:r_vs_ns} for a pivot scale $k_*=0.002\,{\rm Mpc}^{-1}$, together with the newest CMB constraints at the 95\% confidence level arising from a combination of Planck and BICEP/KECK results \cite{BICEP:2021xfz}, {as well as the projected 95\% reach of the Simons Observatory  ($r<0.006$) \cite{SimonsObservatory:2018koc}, BICEP Array ($r<0.006$) \cite{BICEP:2021xfz}, LiteBird ($r<0.002$) \cite{LiteBIRD:2022cnt}, and CMB-S4 ($r<0.001$) \cite{Abazajian:2019eic}}. {Fitting the amplitude of the curvature perturbations inferred from the CMB imposes one relation between the two inflationary parameters $\tilde\lambda_\sigma,\xi_\sigma$; due to this, quantities during inflation can be characterized by a single parameter, which can be chosen as e.g. $\xi_\sigma$ or $r$, as illustrated in Figs.~\ref{fig:r_vs_ns}, \ref{fig:x_vs_xi}.}

In SMASH, inflation ends when $\phi\sim \mathcal{O}(M_P)$, after which the background goes through Hubble-damped oscillations that mimic a radiation fluid. Hence radiation domination starts immediately after inflation, which fixes the number of efolds $N=\Delta \log a$ --where $a$ is the scale factor of the Friedmann-Robertson-Walker (FRW) metric-- between the pivot scale's crossing of the horizon and the end of inflation. This results in the orange band in Fig.~\ref{fig:r_vs_ns}, providing an excellent fit to the data.

For $\lambda_{H\sigma}<0$, the oscillations of the scalar background after inflation allow for an efficient reheating. The reheating temperature was estimated in Ref.~\cite{Ballesteros:2016xej} to be around {$T_{\rm rh}\sim  10^{10}$\,GeV, under the assumption of no exponential growth of Higgs fluctuations. Such an estimate will be improved in this paper by including the Higgs field and its decays in the preheating simulations. The PQ symmetry is restored during reheating, with the axion field acquiring random values, and breaks spontaneously at later times. 
Around the QCD cross over, the axion field becomes massive and starts oscillating, behaving as dark matter in the so-called post-inflationary PQ symmetry breaking scenario, with the correct dark matter abundance reached for $v_\sigma$ between $3.3\times 10^{10}$\,GeV and 
$1.5\times 10^{11}$\,GeV~\cite{Buschmann:2021sdq}.

\section{Parameter ranges and benchmark points.\label{sec:parameters}} 

 {From the above considerations, it should be apparent that the parameter space in SMASH for the bosonic couplings of the field $\sigma$ is significantly constrained by the axion dark matter abundance, the Higgs stability problem, and CMB observations. To recapitulate, the dark matter abundance requires $f_a$ between $10^{10}$ GeV and $10^{11}$ GeV. Higgs stability and perturbativity require $\lambda_{H\sigma}^2/\lambda_{\sigma}$ between $10^{-2}$ and $10^{-1}$. Inflation fixes $\xi_\sigma$ in terms of the effective quartic $\tilde\lambda_\sigma$,  which can be between $\sim10^{-12}$ and $\sim10^{-10}$. From the stability constraints it follows that $\tilde\lambda_{\sigma}$ cannot be very different from $\lambda_\sigma$. Roughly, a given tensor-to-scalar ratio $r$ fixes $\xi_\sigma$ (see Fig.~\ref{fig:r_vs_ns}) which determines $ \tilde\lambda_\sigma\sim\lambda_\sigma$  (see Fig.~\ref{fig:x_vs_xi}). Then the requirement of Higgs stability constrains the values of $\lambda_{H\sigma}$. This, together with the dark matter constraint, means that choosing $r$ roughly specifies all the bosonic couplings of $\sigma$, which then settles the scalar field dynamics which determines GW production.  {While the value of $f_a$ leading to the correct dark matter abundance has a sizable theoretical uncertainty, we note that the GW spectrum is largely insensitive to it, because the dominant GW production happens for field or temperature scales much larger than $f_a$. The latter becomes important at around the PQ phase transition, which can have a small indirect effect on the GWs produced at earlier times as it can impact their redshifting.} A limited freedom remains in the choice of $\lambda_{H\sigma}$ which ensures stability in the perturbative regime; also, it should be noted that the different constraints for the couplings apply in principle at different renormalization group (RG) scales, and there can be subleading RG effects affecting the parameter windows. The fermionic couplings beyond the SM in SMASH, namely the Yukawas of the vector quarks and the right-handed neutrinos, play a secondary role. To start with, they are not directly relevant during inflation or reheating, where bosonic effects dominate. On the other hand, while Yukawa couplings are involved in the production rate of GWs from the thermal plasma, in the weak coupling regime the effect of the new fermions in SMASH will be overwhelmed by that of the SM fields.
 
 From the previous discussion it follows that in order to obtain a good estimate for the possible range of the GW spectrum in SMASH across the available parameter space, it suffices to consider the two extreme values of $r$ that remain compatible with the data. As a consequence of this,}
in order to calculate the spectrum of GWs from SMASH we fixed $f_a=1.2\times10^{11}$ GeV and chose 
for the remaining parameters two extremal benchmark points corresponding to the maximum/minimum values of $r$ within the allowed window $0.036\geq r \geq 0.0037$ between the red dots of Fig.~\ref{fig:r_vs_ns}. We have chosen points satisfying the stability conditions of Ref.~\cite{Ballesteros:2016xej}. 

{\bf Benchmark point 1} (BP1) has $r=0.036,$ $n_s = 0.965,$ $\phi_*=21.4 M_P,$ $\phi_{\rm end}=2.2 M_P,$ $\xi_\sigma(\phi_*)=0.014$, $\tilde\lambda_\sigma(\phi_*)=1.25\times10^{-12},$  where field values are given in the Jordan frame, and $\phi_*$ is the value of the inflaton when the CMB pivot scale  crosses the horizon. The values of the Hubble scale at the crossing and at the end of inflation are ${\cal H}_{\rm inf}(\phi_*)=2.0\times10^{-5}M_P$ and ${\cal H}_{\rm end}=1.8\times10^{-6}M_P$. The number of post-inflationary efolds  assuming radiation domination immediately after inflation is $N_{\rm post}=64.8$. The model's couplings at the  $f_a$ scale are 
$\lambda_\sigma(f_a) =3.0\times 10^{-11},$  $\lambda_{H\sigma}(f_a) =-1.5\times10^{-6},$ $ \lambda_H(f_a) =0.079,$  $Y_{ii}(f_a)=1.2\times10^{-3},$  $y(f_a)=8.5\times10^{-4}$. 

For {\bf benchmark point 2} (BP2) in turn we have: $r=0.0037,$  $n_s =  0.967,$ $\phi_*=8.4 M_P,$ $\phi_{\rm end}=0.76M_P,$ $\xi_\sigma(\phi_*)=1.0,$  $\tilde\lambda_\sigma(\phi_*)=5.3\times10^{-10},$ ${\cal H}_{\rm inf}(\phi_*)=6.5\times10^{-6}M_P,$ ${\cal H}_{\rm end}=2.4\times10^{-6}M_P,$  $N_{\rm post}=65.0$, 
 $\lambda_\sigma(f_a) =4.0\times 10^{-9},$  $\lambda_{H\sigma}(f_a) =-2.4\times10^{-5},$  $\lambda_H(f_a) =0.15,$  $Y_{ii}(f_a)=4.5\times10^{-3},$  $y(f_a)=3.6\times10^{-3}.$ 

 {To ensure accurate  predictions, we calculate them using a renormalization scale of the order of the relevant field or energy scales. For inflation we use $\mu=\phi_*$, while for preheating and thermal processes we use $\mu=f_a$ and $\mu=T$, respectively. The couplings are evolved using the two-loop RG equations of Ref.~\cite{Ballesteros:2016xej}.}

\section{ Primordial GWs from SMASH.} Throughout the previously outlined cosmological history, there are three sources of stochastic GWs. First, one has GWs generated from tensor perturbations during inflation. Secondly, the exponential growth of scalar field fluctuations in the oscillating phase after inflation (preheating) generates a source term for GWs which stops when the fluctuations start to decay. Finally, after reheating is completed and the energy density is dominated by light radiation, thermal fluctuations give rise to new source-terms for GW production, which continues as long as the  fermion and gauge boson abundances remain sizable, i.e. roughly until the breaking of the electroweak symmetry. {We emphasize that the different contributions to the spectrum are not independent. This should be clear from our previous discussion about the GW spectra depending approximately in a single parameter like $r$. Nevertheless, we may additionally point out that the GWs from preheating depend on the initial conditions of the fields and their fluctuations after inflation, while the thermal spectrum depends on the reheating temperature, which is determined by the preheating dynamics. Hence, the calculations of the spectra are not independent and remain tied to each other.}
In the following sections we will go over the contributions from each source to the energy fraction of GWs per logarithmic frequency interval, $\Omega_{\rm GW}(f)$, defined as
$ \Omega_{\rm GW}={\rho_{\rm GW 0}}/{\rho_{c0}}=\int \Omega_{\rm GW}(f){d\log f}$,
where $\rho_{\rm GW 0}$ is the present energy density of GWs and $\rho_{c0}=3{\cal H}^2_0 M_P^2$ the current total energy density. ${\cal H}_0=100 \,h\, {\rm km/s/Mpc}$ is today's Hubble rate, with $h\approx0.68$ \cite{Planck:2018vyg}.

\subsection{ GWs from inflation.}The spectrum of the energy fraction of primordial GWs from inflation is well known and can be approximated as~\cite{Ringwald:2020vei}
\begin{equation}\label{eq:OmegaiGWB}\begin{aligned}
&h^2\, \Omega_{\rm iGWB}(f)
\approx 9.9\times 10^{-17}\times\\
&\times{g_{*\rho}(T_{\rm hc}(f))}
\left[ g_{*s}(T_{\rm hc}(f))\right]^{-\frac{4}{3}} 
\left[\frac{{\cal H}_{\rm inf}(f)}{3\times 10^{13}\,\mathrm{GeV}}\right]^2.
\end{aligned}
\end{equation}
Above, ${\cal H}_{\rm inf}(f)$ is the value of the Hubble constant when the mode corresponding to the frequency $f$ crossed the horizon during inflation, (i.e. when ${\cal H} = k/a=2\pi f a_0/a$, where $a_0$ is the present value of the scale factor, and $k$ is the comoving momentum). {As with the inflationary observables in Figs.~\ref{fig:r_vs_ns},\ref{fig:x_vs_xi}, ${\cal H}_{\rm inf}(f)$ is fully determined once $r$ is chosen; the values in the two benchmark points are illustrated in Fig.~\ref{fig:Hinf}. In eq.~\eqref{eq:OmegaiGWB},} $g_{*\rho}$  and  $g_{*s}$ denote the effective numbers of relativistic degrees of freedom associated with the energy and entropy densities, respectively. {They approach 124.5 at high temperatures, and experience steps at decoupling thresholds, the most important one being associated with the  PQ phase transition.  The temperature of the latter is determined by the scale $f_a$, $T_{\rm PQ}\sim\lambda_\sigma^{1/4}f_a$. $g_{*\rho}$  and  $g_{*s}$ are calculable once the SMASH parameters are fixed, as was done in Ref.~\cite{Ringwald:2020vei}. Finally,}
\begin{equation}\label{eq:Thc}
T_{\rm hc}(f)=\!\frac{10^8 {\rm GeV}f}{1.2\,\rm Hz}\!\!
\, 
\left[\frac{g_{*s}(T_{\rm hc}(f)}{g_{*\rho}(T_{\rm hc}(f))}\right]^{1/2}\!\!\!\! [g_{*s}(T_{\rm hc}(f)]^{-1/6}
\end{equation}
 is the temperature at which the mode re-entered the horizon after reheating. The spectrum of GWs during inflation for the two benchmark points in SMASH is given by the leftmost curves in Fig.~\ref{fig:h2Omega_smash_entire}; the vertical dashed sections  represent the cutoff for frequencies that never exited the horizon during inflation~\footnote{Assuming radiation domination after inflation, the cutoff frequency can be expressed as $f_{\rm knee}^{\Omega_{\rm iGWB}} =7.1\times10^7\,{\rm Hz}\,\left[ \frac{{\cal H}_{\rm end}}{5\times10^{12}\,{\rm GeV}}\right]\left[ \frac{e^{-N_{\rm post}}}{e^{-65}}\right]$.}. At frequencies around 1\,Hz, the spectra feature a 
step  due to the PQ transition which could be detected by DECIGO \cite{Ringwald:2020vei}. { To end the discussion about inflationary GWs, we note that the resulting spectrum in SMASH is similar (up to the subleading features from the PQ phase transition) to that in models of inflation with similar power spectra, such as Starobinsky/Higgs inflation \cite{Starobinsky:1980te,Bezrukov:2007ep}. This degeneracy will be broken by the spectra of preheating and thermal fluctuations, which depend on all the bosonic interactions.}

\begin{figure}[t]
\begin{center}
\includegraphics[width=.9\columnwidth]{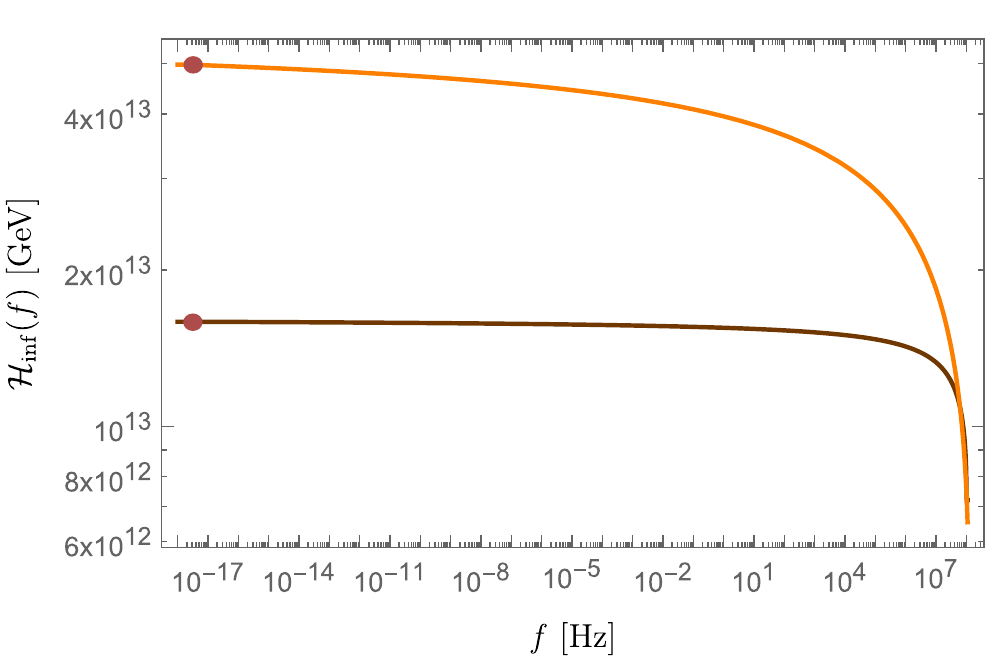}
\end{center}
\vskip-.5cm 
\caption{{Value of the Hubble constant at horizon crossing as a function of the frequency, for BP1 (lighter) and BP2 (darker). The red dots represent the frequency corresponding to the CMB pivot scale of 0.002 $\rm Mpc^{-1}$.}
}
\label{fig:Hinf}
\end{figure}

\subsection{GWs from preheating.} One can estimate the spectrum of GWs in terms of the time-dependent stress-energy tensor of the scalar fields by solving the linearized GW equation in momentum space in a FRW background using Green's function methods \cite{Dufaux:2007pt} (for other approaches, see for example Refs.~\cite{Garcia-Bellido:2007fiu,Easther:2007vj}). This gives~\footnote{We neglect here the differences between $g_{*\rho}$ and $g_{*s}$.}
\begin{align}\label{eq:h2Omega_preheating}\begin{aligned}
 &h^2 \Omega_{\rm pGWB}(f)=\,h^2\Omega_{\rm rad}\times\\
 &\times\!\!\left[\frac{g_{*\rho}(\tau_{\rm rh})}{g_{*\rho}(\tau_0)}\right]^{-1/3}\!\left[\frac{a(\tau_w)}{a(\tau_{\rm rh})}\right]^{1-3w}\!\!\!\!\left.\frac{S_k(\tau_f)}{a(\tau_w)^4\rho(\tau_w)}\right|_{k=2\pi f a_0}\!\!,
 \end{aligned}\end{align}
 with $S_k(\tau_f)$ given by
 \begin{equation}\begin{aligned}
 &S_k(\tau_f)=\,\\
 &\frac{k^3}{2VM_P^2}\!\!\int \!\!d\Omega\!\sum_{m,n}\!\left\{\left|\int_{\tau_i}^{\tau_f}\!\!\!d\tau' \!\cos(k\tau')a(\tau') T^{\rm TT}_{mn}(\tau',\bf k) \right|^2\right.\\
 &\left.+\left|\int_{\tau_i}^{\tau_f}d\tau' \sin(k\tau')a(\tau') T^{\rm TT}_{mn}(\tau',\bf k) \right|^2\right\}.
\end{aligned}\end{equation}
 In the equations above, $h^2\Omega_{\rm rad}=4.2\times10^{-5}$ is the current energy fraction of radiation, $\tau$ denotes conformal time (with current value $\tau_0$ and satisfying $d\tau/dt=1/a$) while $\rho(\tau)$ is the total energy density.
V is the 3D spatial volume, and $\tau_{w}$ is  the moment at which the time-averaged stress-energy tensor reaches a well defined equation of state $p=w\rho$; we expect  $w\approx1/3$. $\tau_{\rm rh}$ denotes the time at which the light particles produced by the inflaton's fragmentation dominate the energy density.   $T^{\rm TT}_{mn}(\tau',\bf k)$ are the Fourier transforms of the spatial components of the transverse-traceless projection of the stress-energy tensor,
\begin{equation}\label{eq:TTsource}\begin{aligned}
 T^{\rm TT}_{mn}(\tau,{\bf k})=\left(\!P_{mp}(\hat{\bf k})P_{nq}(\hat{\bf k})\!-\!\frac{1}{2}P_{mn}(\hat{\bf k})P_{pq}(\hat{\bf k})\right)\times\\
 \times\sum_j \int\frac{d^3{\bf p}}{(2\pi)^{3/2}}\,p_p p_q \varphi_j(\tau,{\bf p}) \varphi_j(\tau,{\bf k-p}).
\end{aligned}\end{equation}
In the equation above, $\hat{\bf k}$ denotes the unit vector in the direction of the 3-momentum $\bf k$, while $P_{mn}({\bf k})=\delta_{mn}-\hat{k}_m\hat{k}_n$ are transverse projectors, and the sum over $j$ runs over all real scalar fields.

\begin{figure}[t]
\begin{center}
\includegraphics[width=.94\columnwidth]{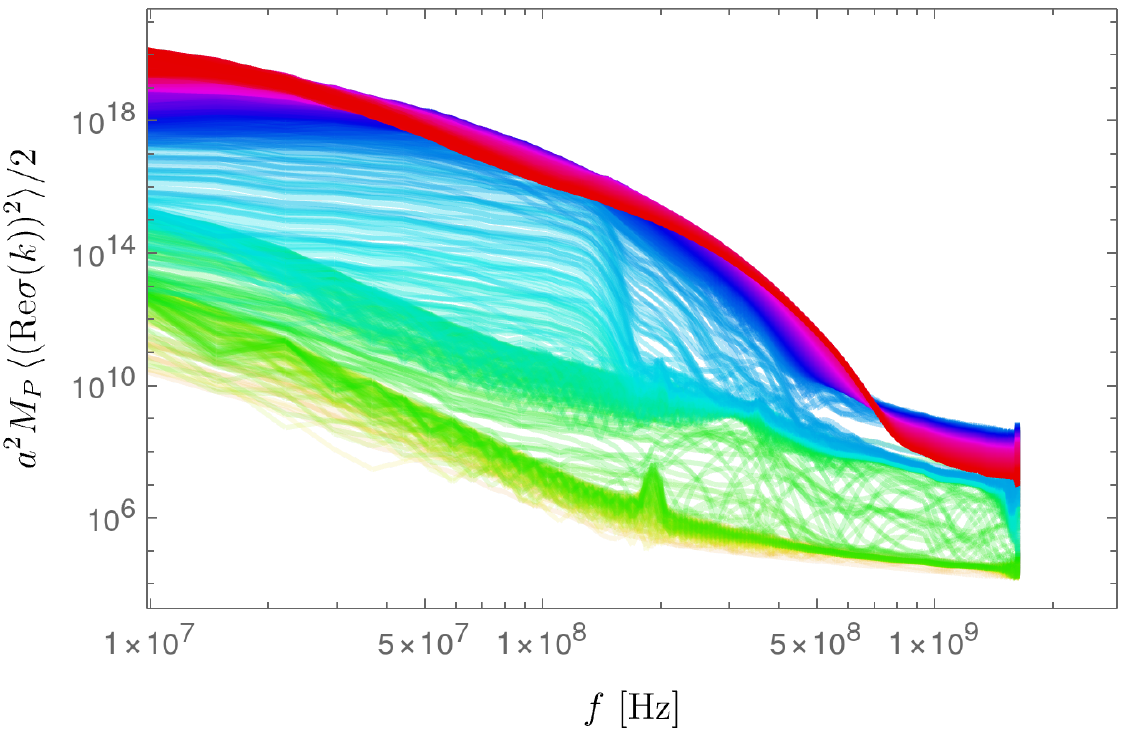}\\
\includegraphics[width=.94\columnwidth]{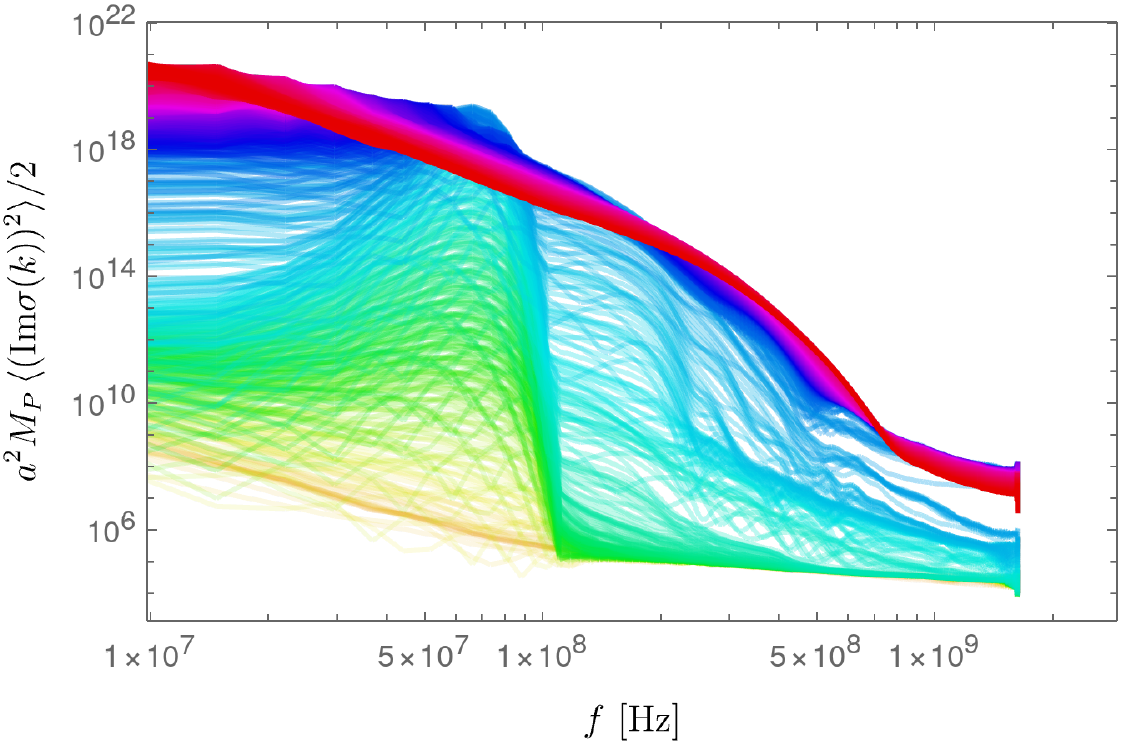}\\
\includegraphics[width=.94\columnwidth]{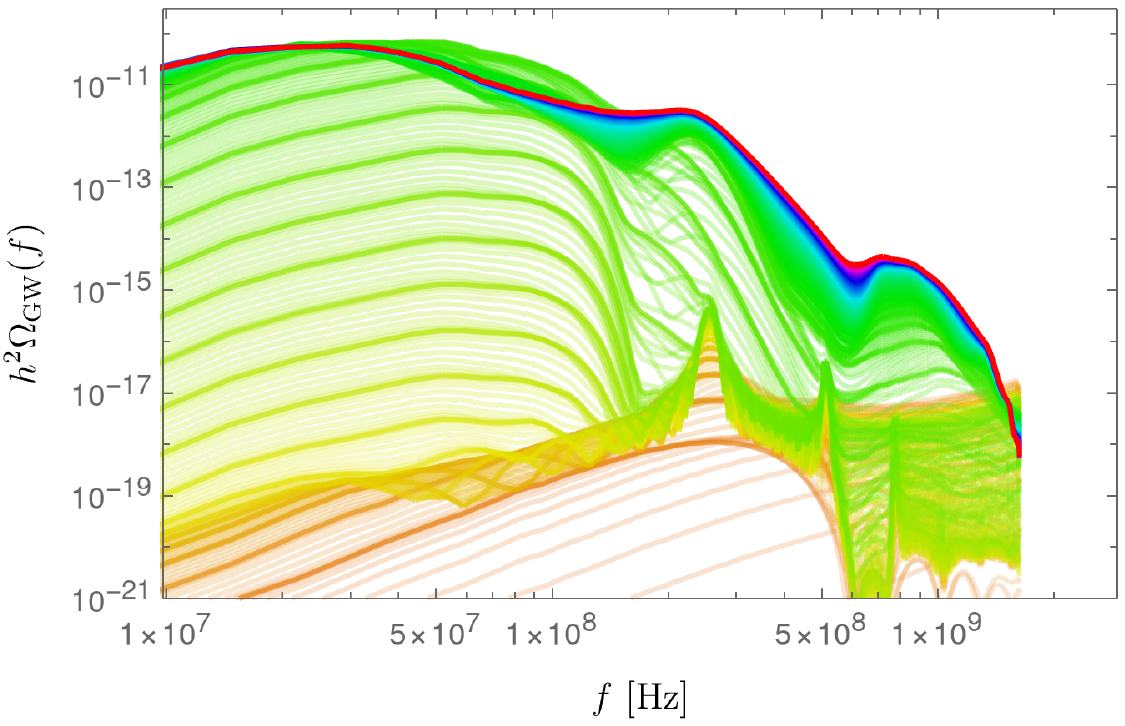}
\end{center}
\vskip-.5cm 
\caption{
Upper/middle plots: Power spectra of ${\rm Re}\,\sigma$/ ${\rm Im}\,\sigma$  for BP1, as a function of today's frequency for subsequent values of the conformal time (earlier times in orange, later times in red). Lower plot: present energy density of GWs for BP1, with the source integrated up to different times, and with a similar color coding. The red lines correspond to the final time of the simulation.
}
\label{fig:hcOmega_smash_preheating}
\end{figure}

As the energy density of GWs is expected to be small, one can neglect their backreaction into the evolution of the scalar fields. 
To compute $h^2\Omega_{\rm pGWB}$ from Eq.~\eqref{eq:h2Omega_preheating} we have resorted to lattice simulations of the evolution of scalar fields in a FRW background, in a similar way as described in Ref.~\cite{Ballesteros:2021bee}.    We have solved the equations for three real scalars --the real and imaginary parts of $\sigma$ and the neutral component of the Higgs-- in lattices with $256^3$ points. The couplings were evaluated at a renormalization scale $\mu=f_a$, 
and we accounted for Higgs decays by including a decay term in the Higgs' equation of motion. We modeled the decay products with a homogeneous relativistic fluid, whose density $\rho_{\rm rad}$ evolves in time ensuring the covariant conservation of the total stress-energy momentum tensor. The scale factor was also evolved in a consistent manner, and the initial conditions were determined  
from the backgrounds and power spectra at the end of inflation, {which were computed by solving the equations for the scalar background and for the linearized fluctuations in momentum space as a function of time. As emphasized earlier, the computation of the GWs during preheating is tied to the results for GWs during inflation.} The computations were carried out with a modified version of {\tt{CLUSTEREASY}} \cite{Felder:2000hq,Felder:2007kat}; see Ref.~\cite{Ballesteros:2021bee} for more details. We took $\tau_w$ as the final time of the simulation and computed $w$ using $w=-1/3(1+2\ddot{a}a/\dot{a}^2)$ (with $\dot{}=d/dt$).  $\tau_{\rm rh}$ was inferred from the results for the energy densities, carrying out extrapolations if necessary. Assuming thermalization in the radiation bath at $\tau_{\rm rh}$, we estimated the reheating temperature as $T_{\rm rh}=(30\,\rho_{\rm rad}(\tau_{\rm rh})/(\pi^2 g_{\star\rho}(T_{\rm rh}))^{1/4}$. By matching the extrapolated Hubble rate to  ${\cal H}_0$, accounting for the late period of matter domination, we estimated $N_{\rm post}=\log a_0/a_{\rm end}$.

The results of the simulations for BP1 are illustrated in Fig.~\ref{fig:hcOmega_smash_preheating}, which shows the power spectra of the fields for different values of time, as well as the present energy density of GWs obtained when integrating the source up to different times. The  spectra of the fields show resonance bands and peaks which are correlated (up to distortions from the  convolution  appearing in Eq.~\eqref{eq:TTsource}) with the peaks in the GW spectrum. 
 The GW spectra for both benchmark points are shown by the middle curves in Fig.~\ref{fig:h2Omega_smash_entire}. Dashed sections represent an extrapolation based on an $f^3$ behaviour for small frequencies \cite{Dufaux:2007pt}, cross-checked with additional simulations.

 For BP1 we infer $w=0.3398$, $N_{\rm post}=64.3,$ $T_{\rm rh}=9.7\times10^{12}$ GeV, $h^2\Omega_{\rm pGWB}=9.5\times10^{-11}$, while for BP2 we obtain $w=0.3334$, $N_{\rm post}=65.0,$ $T_{\rm rh}=2.0\times10^{12}$ GeV, $h^2\Omega_{\rm pGWB}=1.1\times10^{-10}$. The reheating temperatures are significantly higher than the estimates of $T_{\rm rh}\approx 10^{10}$ GeV in Ref.~\cite{Ballesteros:2016xej}, which assumed that no resonant growth of Higgs fluctuations was possible. This is indeed the case during the first oscillations of the background after inflation, but no longer true once the fluctuations of ${\rm Im}\,\sigma$ start becoming amplified, lowering the Higgs mass thanks to the negative portal coupling.  The resulting growth of $\rho_{\rm rad}$ is illustrated in Fig.~\ref{fig:rhorad}.
 
The main features of the GW spectra can be captured by the following parameterizations,
\begin{align}\label{eq:fpeakpreheating}
&f_{\rm peak}^{\rm pGWB}\simeq3.5\times10^{13}\,{\rm Hz}\,\,\hat\kappa\sqrt{\tilde\lambda_\sigma}\,\left[\frac{\phi_{\rm end}}{M_P}\right]\left[\frac{e^{-N_{\rm post}}}{e^{-65}}\right],\\
\nonumber &h^2\Omega_{\rm pGWB}\simeq \frac{1.7\times10^{-7}\alpha}{\hat{\kappa}^2}\!\left[\frac{{\cal H}_{\rm end}}{5\times10^{12}\,{\rm GeV}}\right]^2\!\left[\frac{e^{-4N_{\rm post}}}{e^{-4\cdot65}}\right],
\end{align}
which follow from writing the typical size of field inhomogeneities during the fragmentation process as $\hat{R}=a/(\hat{\kappa} \sqrt{\tilde\lambda_\sigma}\phi_{\rm end} a_{\rm end})$, and estimating the energy fraction in GWs at the onset of fragmentation as $\rho_{\rm GW}(\tau_{\rm frag})/\rho(\tau_{\rm frag})=\alpha(\hat{R}{\cal H}_{\rm frag})^2$ \cite{Dufaux:2007pt}. To arrive to Eq.~\eqref{eq:fpeakpreheating} we further assumed radiation domination (i.e. $\omega=1/3$, as confirmed by the results above) and $\tau_{\rm frag}\approx200/(\sqrt{\lambda_\sigma}\phi_{\rm end}a_{\rm end})$. {The latter is meant to be the time at which fluctuations start being amplified, which can be inferred from Fig.~\ref{fig:rhorad} by identifying the onset of the exponential growth of the density of the SM radiation bath, which is driven by Higgs fluctuations}. Eqs.~\eqref{eq:fpeakpreheating} can fit the peak frequency and total energy fraction in BP1/BP2 with $\hat{\kappa}=0.05$/0.08 and $\alpha=1\times10^{-5}$/$3\times10^{-4}$. {Rather than free constants, $\hat{\kappa}$ and $\alpha$ are deduced from the simulations and correspond to a simplified parameterization of the results. The rest of the parameters in Eq.~\eqref{eq:fpeakpreheating} are fixed by the inflationary dynamics and are determined once $r$ is fixed, as illustrated in Figs.~\ref{fig:r_vs_ns}, \ref{fig:x_vs_xi}. The fact that somewhat different values of $\hat\kappa$ and $\alpha$ are deduced from the simulations for BP1 and BP2 is not entirely surprising because, while $\phi_{\rm end}$, ${\cal H}_{\rm end}$ and $N_{\rm post}$ are largely insensitive to the Higgs portal coupling $\lambda_{H\sigma}$, the latter should play a role in determining the production of Higgs fluctuations, which affects inflaton fragmentation. Hence we expect the effective parameters $\alpha,\hat\kappa$ to depend on $\lambda_{H\sigma}$, which again is constrained by stability requirements once $r$ is fixed.}

\begin{figure}[t]
\begin{center}
\includegraphics[width=.9\columnwidth]{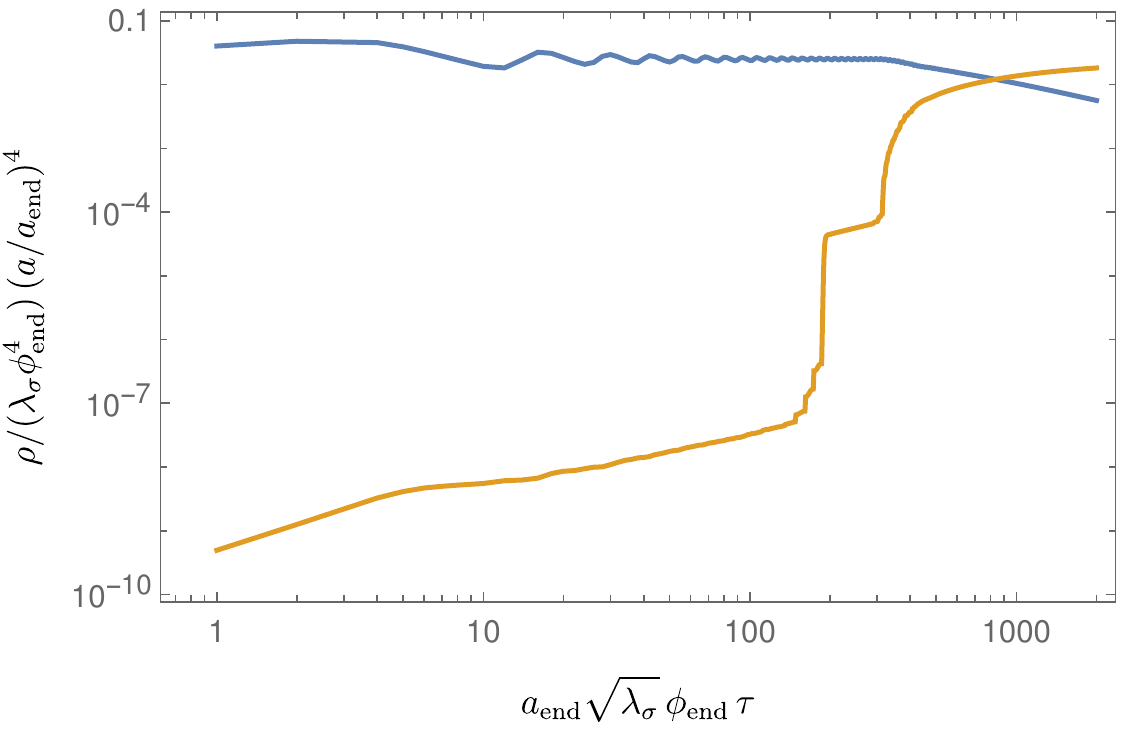}
\end{center}
\vskip-.5cm 
\caption{Evolution of the mean energy densities of the scalars (blue) and radiation bath (orange) for BP1, giving $\tau_{\rm rh}=835/(\sqrt{\lambda_\sigma}\phi_{\rm end}a_{\rm end})$ captured within the simulation.
}
\label{fig:rhorad}
\end{figure}

\subsection{ GWs from thermal fluctuations.} The Cosmic Gravitational Microwave Background (CGMB) arising from thermal fluctuations in the plasma during radiation domination has been studied in Refs.~\cite{Ghiglieri:2015nfa,Ghiglieri:2020mhm,Ringwald:2020ist}\footnote{See also \cite{Klose:2022knn} for an application to a model of axion-inflation.}, giving
\begin{equation}\label{eq:Omega_CGMB_analytic_expression}\begin{aligned}
&{{h^2}}\,\Omega_{\rm CGMB}  (f)
\approx  
4.0\times 10^{-12} 
\!
\left[\frac{T_{\rm rh}}{M_P}\right]\!\left[\frac{g_{*s}(T_{\rm rh})}{106.75}\right]^{-5/6}\!\!\!\!\!\!\times\\
&\times
\left[ \frac{f}{\rm GHz}\right]^3
\hat\eta\left(
T_{\rm rh},2\pi \,
\left[\frac{g_{*s}(T_{\rm rh})}{3.9}\right]^{1/3}\,
\frac{f}{T_0}
\right)
 \;,
\end{aligned}\end{equation}
where $T_{\rm rh}$
and $T_0$  denote  the maximum temperature of the plasma after it starts dominating the energy density 
and the current CMB temperature, respectively. 
The function $\hat\eta$ is only known for low and high values of $k/T$. For the latter, $\hat\eta$ 
has been computed for the SM with full leading order precision in Ref.~\cite{Ghiglieri:2020mhm}, and the result was generalized to arbitrary models in Ref.~\cite{Ringwald:2020ist}. The ensuing spectrum has an amplitude scaling with $T_{\rm rh}$, and peaking at a frequency of the order of $80 (106.75/(g_{*s}(T_{\rm rh}))^{1/3} $ GHz. Hence a precise measurement of the CGMB could inform us of the temperature and degrees of freedom of the primordial plasma. {The main dependence of the thermal spectrum on the SMASH parameters is through the value of $T_{\rm rh}$, which is fixed by the scalar dynamics during preheating and is thus associated with $\lambda_\sigma,\lambda_{H\sigma}$, which as elaborated before are constrained  once $r$ is fixed. The function $\hat\eta$ is independent of the scalar couplings (as scalar interactions do not produce GWs at leading order) and in principle depends on the SMASH Yukawa couplings. However, as mentioned before in the weak coupling regime their effect remains subleading with respect to that of SM Yukawas.} Using the values of $T_{\rm rh}$ inferred from the preheating simulations, the thermal spectrum for the two benchmarks is shown by the rightmost curves in Fig.~\ref{fig:h2Omega_smash_entire}; the dashed lines interpolate between the results for low/high $k/T$.

\section{Discussion} The collected spectra of GWs in SMASH are shown in Fig.~\ref{fig:h2Omega_smash_entire}. {As argued in Section~\ref{sec:parameters}, given how CMB and stability constraints allow to limit the choices for all scalar couplings of $\sigma$ once the tensor-to-scalar ratio $r$ is chosen, we expect the two benchmark spectra for maximal and minimal $r$ to provide a very good estimate of the range of results that can be obtained in the full parameter space. From the outcome it} can be seen that inflaton fragmentation gives the largest emission of GWs in the frequency range  between 
 $\sim 10^{5\div 6}$\,Hz and $10^{9\div 10}$\,Hz, while the inflationary GWs and the thermal GWs dominate below and above this frequency window, respectively. The peaks of the preheating and thermal spectra are well separated, and the three different components in the spectrum could be disentangled from each other if experiments were to reach the required sensitivities. A hypothetical measurement of the GW spectrum between $\sim 1$ Hz and $100$ GHz could potentially determine the Hubble scale during inflation --which enters $\Omega_{\rm iGWB}$-- the scale of inflaton fragmentation after inflation --related to $f^{\rm pGWB}_{\rm peak}$-- and finally the  maximum temperature and the number of relativistic degrees of freedom of the hot Big Bang plasma, which fix the amplitude and peak of $\Omega_{\rm CGMB}$.  This could provide an unprecedented window into the physics of the very early universe. {In perturbative realizations of SMASH with a stable scalar potential, all the previous physical quantities can be related to a single parameter, the tensor-to-scalar ratio $r$, up to RG running effects and a limited freedom in the choice of the portal coupling $\lambda_{H\sigma}$ ensuring stability. This shows that the shape of the spectrum in SMASH is significantly constrained, which opens new avenues for the possibility of falsifying the model in the case of hypothetical future measurements of the high-frequency spectrum.}

 We expect the main features of the spectrum of Fig.~\ref{fig:h2Omega_smash_entire} to be generic and representative of a wide class of models featuring inflation and preheating followed by radiation domination. {As mentioned in the introduction, our} choice of model can be considered as  a conservative benchmark, as it does not feature GWs sourced by first-order phase transitions, or an appreciable fraction of GWs from cosmic strings~\footnote{Although cosmic strings are formed in SMASH during the PQ phase transition, their associated energy scale fixed by $f_a\sim10^{11}\,{\rm GeV}$ is too low to produce an appreciable GW signal.}.

 In Fig.~\ref{fig:hcOmega_smash_entire} we show the dimensionless strain $h_c(f)=\sqrt{3{\cal H}_0^2\Omega_{\rm GW}(f)/(2\pi^2)}/f$ predicted in SMASH, confronted with current and projected experimental limits~\cite{LIGOScientific:2016wof,Seto:2001qf,Punturo:2010zz,LISA:2017pwj,BBO_proposal,Aggarwal:2020umq,Holometer:2016qoh,Goryachev:2014yra,Akutsu:2008qv,Domcke:2020yzq,Ito:2019wcb,Ito:2020wxi,Ejlli:2019bqj,Domcke:2022rgu,Ringwald:2020ist,Berlin:2021txa,Schmitz:Zenodo,Herman:2022fau} as well as indirect dark radiation constraints~\cite{Pagano:2015hma,Clarke:2020bil}, together with the dark radiation limit that would correspond to the theoretical uncertainty in the number of effective neutrino species~\cite{Ghiglieri:2020mhm}.

In regards to the prospects for observational detection, a potential timeline could be the following. First, the upcoming generation of CMB experiments such as the BICEP Array \cite{Hui:2018cvg}, CMB-S4~\cite{Abazajian:2019eic}, LiteBIRD~\cite{LiteBIRD:2022cnt}, and the Simons Observatory~\cite{SimonsObservatory:2018koc} has the capability to detect the non-zero  tensor-to-scalar ratio $r$ predicted by SMASH (cf.~Fig.~\ref{fig:r_vs_ns}). Given a positive measurement, future spaceborne GW interferometers such as BBO~\cite{BBO_proposal} or DECIGO~\cite{Seto:2001qf} would be sensitive to $\Omega_{\rm iGWB}$ (cf.~Fig.~\ref{fig:hcOmega_smash_entire}), while Ultimate DECIGO~\cite{Kuroyanagi:2014qza} could potentially detect the step-like feature in the spectrum 
at around 1 Hz due to the PQ phase transition~\cite{Ringwald:2020vei}. The frequency of the step could be cross-checked with the indirect determination 
of $f_a$ resulting from the potential measurement of the axion mass, $m_a \simeq 
57\, {\rm \mu eV}   \left({10^{11}\,{\rm GeV}}/{f_a}\right)$,  by axion dark matter direct detection experiments sensitive in the mass region favored in the post-inflationary PQ symmetry breaking scenario predicted by SMASH, $m_a > 28(2)\,{\rm \mu eV}$~\cite{Borsanyi:2016ksw}, 
 such as for example MADMAX~\cite{MADMAX:2019pub}. 
Probing the waves generated by preheating and thermal effects requires much progress in the detection of ultra high frequency GWs (cf.~Fig.~\ref{fig:hcOmega_smash_entire}). Such efforts are very well motivated by the prospect to probe physics shortly after inflation, and a worldwide initiative towards this goal has already started~\cite{Aggarwal:2020olq}.

\section*{Acknowledgments.} We would like to thank Yvette Welling for discussions in the early stage of this project. 
AR acknowledges support by the Deutsche Forschungsgemeinschaft (DFG, German Research Foundation) under Germany's Excellence Strategy -- EXC 2121 ``Quantum Universe'' -- 390833306. CT acknowledges financial support by the DFG through SFB 1258 and the ORIGINS cluster of excellence.
\newpage
\begin{widetext}
\onecolumngrid
\begin{figure}[h]
\begin{center}
\includegraphics[width=.95\textwidth]{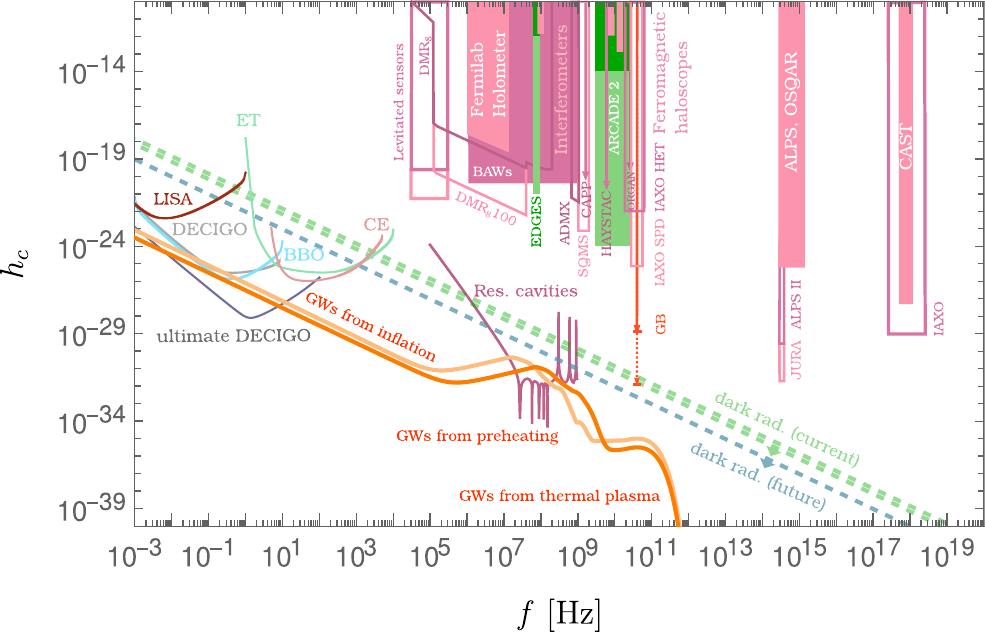}
\end{center}
\vskip-.5cm 
\caption{ Characteristic amplitude of primordial GWs in SMASH (orange) compared to present  (shaded areas) and projected limits (colored solid lines)~\cite{LIGOScientific:2016wof,Seto:2001qf,Punturo:2010zz,LISA:2017pwj,BBO_proposal,Aggarwal:2020umq,Holometer:2016qoh,Goryachev:2014yra,Akutsu:2008qv,Domcke:2020yzq,Ito:2019wcb,Ito:2020wxi,Ejlli:2019bqj,Domcke:2022rgu,Ringwald:2020ist,Berlin:2021txa,Schmitz:Zenodo,Herman:2022fau}. Indirect dark radiation constraints~\cite{Pagano:2015hma,Clarke:2020bil,Ghiglieri:2020mhm} are shown with dashed lines. Abbreviations, BAWs: bulk
acoustic wave devices, SPD: single photon detection, HET: heterodyne detection, Res.: resonant, GB: Gaussian beam, and rad.:
 radiation.
}
\label{fig:hcOmega_smash_entire}
\end{figure}
\end{widetext}
\twocolumngrid
\appendix
\section{Equations solved in the lattice simulations}
Here we provide some details on the equations implemented in our lattice simulations. The dynamical variables are 3 real scalar fields, the homogeneous  density $\rho_{\rm rad}(t)$ from the Higgs decay products, and the scale factor $a(t)$. Denoting the canonically normalized real fields as $\phi=\sqrt{2}\{{\rm Re}H^0,{\rm Re}\sigma,{\rm Im}\sigma\}$, the equations can be written as
\begin{align}\label{eq:eomlattice}\begin{aligned}
&\ddot\phi_n+3\frac{\dot a}{a}\dot\phi_n-\frac{1}{a^2}\vec{\nabla}^2\phi_n+\frac{\partial V(\phi_m)}{\partial\phi_n}+\Gamma_n\dot{\phi}_n=0, \,n\leq3,\\
&\dot{\rho}_{\rm rad}+4 {\frac{\dot a}{a}} \rho_{\rm rad} - \Gamma_1 \dot{h}^2=0,
\end{aligned}\end{align}\begin{align*}
\begin{aligned}
&3M^2_P\left(\frac{\dot a}{a}\right)^2=\rho_{\rm SM}+V_J+\frac{1}{2}\sum_n\dot{\phi}_n^2+\frac{1}{2a^2}\sum_n(\vec{\nabla}\phi_n)^2\,.
\end{aligned}\end{align*}
Above, dots are time derivatives, and $\vec{\nabla}$ denote spatial gradients.
The $\Gamma_n$ are meant to be decay rates. We consider only Higgs decays, i.e. $\Gamma_2=\Gamma_3=0$, while for $\Gamma_1$ we take the perturbative Higgs decay rate,
\begin{align}
\label{eq:Gammah}\begin{aligned}
 \Gamma_1=& \,\Gamma_{h\rightarrow t\bar t}+\Gamma_{h\rightarrow b\bar b}+\Gamma_{h\rightarrow W^+W^-}+\Gamma_{h\rightarrow ZZ},\\
 \Gamma_{h\rightarrow t\bar t}=&\,\frac{3y_t^2}{16\pi}m_h\left(1-\frac{4m^2_t}{m^2_h}\right)^{3/2},\\
  \Gamma_{h\rightarrow b\bar b}=&\,\frac{3y_b^2}{16\pi}m_h\left(1-\frac{4m^2_b}{m^2_h}\right)^{3/2},\\
 \Gamma_{h\rightarrow ZZ}=&\,\frac{g^2}{128\pi}\frac{m^3_h}{m^2_W}\sqrt{1-x_Z}\left(1-x_Z+\frac{3}{4}x
 ^2_Z\right),
 \end{aligned}\end{align}\begin{align*}
\begin{aligned}
 \Gamma_{h\rightarrow W^+W^-}=&\,\frac{g^2}{64\pi}\frac{m^3_h}{m^2_W}\sqrt{1-x_W}\left(1-x_W+\frac{3}{4}x
 ^2_W\right),
\end{aligned}\end{align*}
where
\begin{align}\label{eq:xs}\begin{aligned}
 x_{Z/W}=&\,\frac{4m^2_{Z/W}}{m^2_h},& m^2_W=&\,\frac{g^2 \langle h^2\rangle}{4}, \\ m^2_Z=&
 ,\frac{(g^2+{g'}^2)\langle h^2\rangle}{4}.
\end{aligned}\end{align}
 We substitute the squares of the Higgs mass and vacuum expectation value with the lattice averages of $\partial^2 V/\partial h^2$ and $h^2$ at a given time. Since for $\langle h^2\rangle\rightarrow0$ there is no symmetry breaking  and the computation of the decay rates assuming three massive gauge boson polarizations breaks down, the decay rates into $W,Z$ diverge. Nevertheless, during the time evolution, the fast growth of Higgs fluctuations quickly gives $\langle h^2\rangle\neq0$. For numerical stability at early times we only  activate the $W,Z$ decay channels for $x_W>10^{-3}$.}

\newpage

\bibliography{smashgwBIB}

\end{document}